\documentclass[amsmath,preprint,tightenlines,aps,prd,showpacs,nofootinbib]{revtex4}

\def\be{\begin{equation}}
\def\ee{\end{equation}}
\def\bea{\begin{eqnarray}}
\def\eea{\end{eqnarray}}
\def\bml{\begin{subequations}}
\def\blea{\bml\begin{eqnarray}}
\def\elea{\end{eqnarray}\end{subequations}}

\begin{document}

\title{Geodesics in the static Mallett spacetime}

\author{Ken D. Olum}
\email{kdo@cosmos.phy.tufts.edu}
\affiliation{Institute of
Cosmology, Department of Physics and Astronomy, Tufts University,
Medford, MA  02155}

\begin{abstract}

Mallett has exhibited a cylindrically symmetric spacetime containing
closed timelike curves produced by a light beam circulating around a
line singularity.  I analyze the static version of this spacetime
obtained by setting the intensity of the light to zero.  Some null
geodesics can escape to infinity, but all timelike geodesics in this
spacetime originate and terminate at the singularity.  Freely falling
matter originally at rest quickly attains relativistic velocity inward
and is destroyed at the singularity.

\end{abstract}


\maketitle

\section{Introduction}

A few years ago, Mallett \cite{Mallett} exhibited a spacetime
involving light rays circulating around an infinitely long cylinder in
the manner of the current in a solenoid.  He showed that this
spacetime contains timelike paths that move purely azimuthally and so
close on themselves after going once around the light cylinder.
Although Ref.~\cite{Mallett} discusses photonic crystals as a possible
means of keeping the light in its circular path, in fact so no such
mechanism is necessary.  Rather, as shown in Ref.~\cite{Olum:2004kz},
the light rays travel along geodesics orbiting a central line
singularity.

Ref.~\cite{Mallett} does not discuss the singularity, but in
Ref.~\cite{MallettBook}, Mallett says that he introduced it to confine
the light rays to the cylinder in a way leading to simpler
calculations, and that the spacetime of Ref.~\cite{Mallett} is thus
the combination of effects due to the circulating light and those due
to the singular source.  However, Ref.~\cite{Olum:2004kz} argues that
this means that the spacetime of Ref.~\cite{Mallett} is unlike the
spacetime one would get by introducing circulating light into a flat
background.  To further investigate this question, we will examine the
static spacetime resulting from the line singularity without any
circulating light.  This spacetime does not have causality violation
\cite{MallettBook}, but it does have unusual properties, as I will
discuss below.

The full Mallett spacetime cannot be constructed, because the cylinder
and the singularity are infinitely long.  But with advanced technology
that would permit us to produce singularities, we could presumably set
out to construct a finite approximation to this spacetime.  We could
first produce the background and then introduce a very intense beam of
light in an effort to produce closed timelike paths.  But even the
background is quite problematic.  As I show below, the orbiting
geodesics discovered by Mallett are the only complete geodesics in
this spacetime.  Every other null geodesic perpendicular to the
singular line either originates or terminates at the singularity;
every null geodesic not perpendicular to the singular line both
originates and terminates there, as does every timelike geodesic.

\section{Geodesics}

In Ref.~\cite{Mallett}, the intensity of the circulating light beam is
a free parameter $\epsilon$.  By setting $\epsilon = 0$, we get the
static spacetime without the light.  The metric then becomes
\be
ds^2 = - (\rho/\alpha) dt^2+\rho\alpha
d\phi^2+\sqrt{\alpha/\rho}(d\rho^2+dz^2)\,.
\ee
I have used here the metric signature $(-+++)$, opposite to that of
Ref.~\cite{Mallett}.  The parameter $\alpha$ is a constant with the
dimensions of length, presumably related to the radius of the light cylinder.
To simplify the computation, we can go to units in which $\alpha = 1$, giving
\be
ds^2 = \rho(-dt^2+d\phi^2)+\rho^{-1/2}(d\rho^2+dz^2)\,.
\ee

It is straightforward to compute the connection and write out the
components of the geodesic
equation.  For a geodesic $x(\lambda)$, we find
\bea
\ddot t + \frac{1}{\rho}\dot\rho\dot t &= & 0\,,\label{eqn:tddot}\\
\ddot\phi + \frac{1}{\rho}\dot\rho\dot\phi &= & 0\,,\label{eqn:phiddot}\\
\ddot z - \frac{1}{2\rho}\dot\rho\dot z &= & 0\,,\label{eqn:zddot}\\
\ddot \rho + \frac{1}{4\rho}\left(\dot z^2-\dot \rho^2\right)
+\frac{\sqrt{\rho}}{2}\left(\dot t^2-\dot\phi^2\right)&=&0\,,\label{eqn:rhoddot}
\eea
where a dot denotes differentiation with respect to $\lambda$.
From Eq.~(\ref{eqn:tddot}--\ref{eqn:zddot}), we can write 3 constants
of the motion,
\bea
T &=& \rho\dot t = \text{constant}\\
L &=& \rho\dot\phi = \text{constant}\\
Z &=& \rho^{-1/2}\dot z = \text{constant}
\eea
which make Eq.~(\ref{eqn:rhoddot})
\be\label{eqn:rho2}
\ddot \rho = \frac{1}{4\rho}\dot \rho^2
+\frac{1}{2\rho^{3/2}}\left(L^2-T^2\right) -\frac{1}{4}Z^2\,.
\ee

Now define
\be\label{eqn:V}
V = -\dot x^a\dot x_a = \rho (\dot t^2-\dot \phi^2)
-\rho^{-1/2}(\dot\rho^2 + \dot z^2)
= - \rho^{-1/2}\dot\rho^2 +\rho^{-1}(T^2-L^2) - \rho^{1/2} Z^2\,.
\ee
For a timelike geodesic, we can parameterize with proper time, making
$V = 1$.  For a null geodesic, $V = 0$.   Equation (\ref{eqn:rho2})
ensures that the derivative of Eq.~(\ref{eqn:V}) vanishes.

Now we are in a position to analyze the geodesics.  First use
Eq.~(\ref{eqn:V}) to eliminate $T$ and $L$ in Eq.~(\ref{eqn:rho2}).
We find
\be\label{eqn:rho3}
\ddot \rho = -\frac{1}{4\rho}\dot \rho^2
-\frac{3}{4}Z^2-\frac{1}{2\sqrt{\rho}} V\le 0
\ee
This is utterly unlike the normal motion in polar coordinates.  Since
$\ddot\rho\le0$, there is no geodesic which passes by the singularity,
attaining a minimum distance and then traveling away again.  If a
geodesic is directed inward ($\dot\rho<0$), it eventually
collides with the singularity.

What about an outward-directed geodesic?  Could it escape to infinity?
Consider the right hand side of Eq.~(\ref{eqn:V}) as $\rho\to\infty$.
If $Z\ne0$, it goes to $-\infty$, clearly impossible.  If $Z=0$, it
goes to 0, which is possible only if $V=0$.  So no timelike geodesic,
nor any geodesic that moves in the $z$ direction, can escape to
infinity.

Now consider Eq.~(\ref{eqn:rho3}).  If $Z\ne0$, then $\ddot\rho$ is
bounded by a negative number, so eventually $\dot\rho$ becomes
negative: the geodesic turns around and moves inward to the
singularity.  Since we know that $\rho$ cannot grow to $\infty$, the
same argument applies for any timelike ($V = 1$) geodesic.  Thus all
timelike geodesics and all geodesics with any motion in $z$ originate
and terminate in the singularity.

What about null geodesics perpendicular to the singularity, which have
$Z=V=0$?  In that case, if we set $\dot\rho=0$ we find $\ddot\rho=0$,
so a photon can orbit at any fixed $\rho$.  This is the path found by
Mallett.  More generally, Eq.~(\ref{eqn:V}) becomes
$\dot\rho^2=\rho^{-1/2}(T^2-L^2)$.  If $T^2 < L^2$ there are no
solutions.  If $T^2= L^2$, we get the circular orbit. If
$T^2> L^2$ the general solution is
\be
\rho = [c(\lambda -\lambda_0)]^{4/5}
\ee
where
\be
c = \pm\frac{5}{4}\sqrt{T^2-L^2}\,.
\ee
By choice of parameterization of the null geodesic, we can eliminate
both $\lambda_0$ and $c$, resulting in simply
\be\label{eqn:null}
\rho = \pm\lambda^{4/5}
\ee
The upper sign in Eq.~(\ref{eqn:null}) represents a geodesic starting
from the singularity at parameter $\lambda = 0$ and going out to
infinity.  The lower represents a geodesic coming in from infinity and
terminating at the singularity at $\lambda=0$.

If $L=0$, this geodesic is radial.  Otherwise, it winds around the
singularity.  The angular velocity of the outgoing geodesic is
$\dot\phi = L\lambda^{-4/5}$.  Thus
\be
\phi(\lambda)-\phi (0) = 5L\lambda^{1/5}\,.
\ee
The geodesic winds a finite number of times near the singularity, and
an infinite number of times as it goes out to infinity.

Now let us return to timelike geodesics and consider the fate of a
particle initially at rest at some position $\rho_0$.  It will always
have $L = Z = 0$, so Eq.~(\ref{eqn:V}) becomes
\be
1= -\rho^{-1/2}\dot\rho^2 +\rho^{-1}T^2
\ee
At $\rho_0$, $\dot\rho=0$, so $T^2=\rho_0$, and we have
\be
\dot\rho^2=\rho_0\rho^{-1/2}-\rho^{1/2}
\ee
Relabeling the proper time parameter $\tau$, and choosing the
inward-going path in the future of the initial time, we find
\be
\frac{d\tau}{d\rho} = -\frac{\rho^{1/4}}{\sqrt{\rho_0-\rho}}\,.
\ee
If we start at rest at $\tau = 0$, integration gives $\rho=0$ at
\be
\tau = B\left(\frac{1}{2}, \frac{5}{4}\right) \rho_0^{3/4}\,,
\ee
where $B$ is the Euler beta function; $B(1/2, 5/4) = 1.748\ldots.$

The proper distance from the singularity to the position with $\rho
=\rho_0$ is given by
\be
R =\int_0^{\rho_0} \rho^{-1/4} d\rho = \frac{4}{3}\rho_0^{3/4}
\ee
We conclude that a particle initially at rest at proper distance $R$
will be destroyed at the singularity after proper time $(3/4)B(1/2,
5/4) R \approx 1.3 R$.

\section{Conclusion}

I have shown that nearly every geodesic in the Mallett spacetime
originates and terminates at the singular line.  The only exceptions
are null geodesics perpendicular to the line singularity.  Such
geodesics with any outward motion originate at the singularity and go
to infinity, those with inward motion originate at infinity and
terminate at the singularity, and those with no radial motion orbit at
fixed radius.

Even a single causal path going from a singularity to some point $p$
makes it impossible to predict what will happen at $p$, because the
information coming from the singularity cannot be known.  But one
might perhaps finesse this issue if $p$ is far from the singularity
and its influence is diluted by distance.  But in the static Mallett
spacetime considered here, from any point the singularity fills the
entire sky except for an infinitesimally thin strip, so the loss of
predictability is much more severe. 

Furthermore, all timelike geodesics terminate in the singularity, so
any freely falling object will eventually reach the singular line.  If
the object begins at rest, its remaining proper lifetime is of order
its proper distance to the singularity.  It therefore appears that any
attempt to build a ``time machine'' along the lines described by
Mallett would have a very unfortunate effect on nearby objects.

\section*{Acknowledgments}

I would like to thank Allen Everett and Tom Roman for their help.
This work was supported in part by the National Science Foundation
under grant 0855447.

\end{document}